\documentclass[a4paper,11pt]{article}
\pdfoutput=1 

\usepackage{ifthen} 
\newboolean{uprightparticles}
\setboolean{uprightparticles}{false} 

\newcommand{\BtoDK}{\ensuremath{\Bpm\to\PD\Kpm}\xspace}

\newcommand{\xpm}{\ensuremath{x_{\pm}}\xspace}
\newcommand{\ypm}{\ensuremath{y_{\pm}}\xspace}
\newcommand{\xp}{\ensuremath{x_{+}}\xspace}
\newcommand{\yp}{\ensuremath{y_{+}}\xspace}
\newcommand{\xm}{\ensuremath{x_{-}}\xspace}
\newcommand{\ym}{\ensuremath{y_{-}}\xspace}
\def\xy {\ensuremath{x_{\pm}}\text{ and }\xspace \ensuremath{y_{\pm}}\xspace}

\usepackage{xspace} 
\usepackage{upgreek}

\def\lhcb {\mbox{LHCb}\xspace}

\def\babar  {\mbox{BaBar}\xspace}
\def\belle  {\mbox{Belle}\xspace}
\def\cleo   {\mbox{CLEO}\xspace}

\def\MagUp {\mbox{\em Mag\kern -0.05em Up}\xspace}

\ifthenelse{\boolean{uprightparticles}}%
{

 \def\Ppi         {\ensuremath{\uppi}\xspace}

 \def\PDelta      {\ensuremath{\Delta}\xspace}                 
 \def\PXi      {\ensuremath{\Xi}\xspace}                 
 \def\PLambda      {\ensuremath{\Lambda}\xspace}                 
 \def\PSigma      {\ensuremath{\Sigma}\xspace}                 
 \def\POmega      {\ensuremath{\Omega}\xspace}                 
 \def\PUpsilon      {\ensuremath{\Upsilon}\xspace}                 
 
 \def\PB      {\ensuremath{\mathrm{B}}\xspace}                 
                  
 \def\PD      {\ensuremath{\mathrm{D}}\xspace}

 \def\PK      {\ensuremath{\mathrm{K}}\xspace}

 \def\Pi      {\ensuremath{\mathrm{i}}\xspace}

}
{

 \def\Ppi         {\ensuremath{\pi}\xspace}

 \mathchardef\PDelta="7101
 \mathchardef\PXi="7104
 \mathchardef\PLambda="7103
 \mathchardef\PSigma="7106
 \mathchardef\POmega="710A
 \mathchardef\PUpsilon="7107
                  
 \def\PB      {\ensuremath{B}\xspace}                 
                  
 \def\PD      {\ensuremath{D}\xspace}

 \def\PK      {\ensuremath{K}\xspace}

 \def\Pi      {\ensuremath{i}\xspace}

}

\makeatletter
\ifcase \@ptsize \relax
  \newcommand{\miniscule}{\@setfontsize\miniscule{4}{5}}
\or
  \newcommand{\miniscule}{\@setfontsize\miniscule{5}{6}}
\or
  \newcommand{\miniscule}{\@setfontsize\miniscule{5}{6}}
\fi
\makeatother

\DeclareRobustCommand{\optbar}[1]{\shortstack{{\miniscule (\rule[.5ex]{1.25em}{.18mm})}
  \\ [-.7ex] $#1$}}

\def\g      {{\ensuremath{\Pgamma}}\xspace}

\def\pion   {{\ensuremath{\Ppi}}\xspace}
\def\piz    {{\ensuremath{\pion^0}}\xspace}

\def\pip    {{\ensuremath{\pion^+}}\xspace}
\def\pim    {{\ensuremath{\pion^-}}\xspace}
\def\pipm   {{\ensuremath{\pion^\pm}}\xspace}

\def\kaon    {{\ensuremath{\PK}}\xspace}
  \def\Kbar    {{\kern 0.2em\overline{\kern -0.2em \PK}{}}\xspace}

\def\KorKbar    {\kern 0.18em\optbar{\kern -0.18em K}{}\xspace}
\def\Kz      {{\ensuremath{\kaon^0}}\xspace}
\def\Kzb     {{\ensuremath{\Kbar{}^0}}\xspace}
\def\Kp      {{\ensuremath{\kaon^+}}\xspace}
\def\Km      {{\ensuremath{\kaon^-}}\xspace}
\def\Kpm     {{\ensuremath{\kaon^\pm}}\xspace}

\def\KS      {{\ensuremath{\kaon^0_{\mathrm{ \scriptscriptstyle S}}}}\xspace}
\def\KL      {{\ensuremath{\kaon^0_{\mathrm{ \scriptscriptstyle L}}}}\xspace}
\def\Kstarz  {{\ensuremath{\kaon^{*0}}}\xspace}

\def\Kstarp  {{\ensuremath{\kaon^{*+}}}\xspace}
\def\Kstarm  {{\ensuremath{\kaon^{*-}}}\xspace}
\def\Kstarpm {{\ensuremath{\kaon^{*\pm}}}\xspace}

\def\Dbar    {{\kern 0.18em\overline{\kern -0.18em \PD}{}}\xspace}
\def\D       {{\ensuremath{\PD}}\xspace}
\def\Db      {{\ensuremath{\Dbar}}\xspace}
\def\DorDbar    {\kern 0.18em\optbar{\kern -0.18em D}{}\xspace}
\def\Dz      {{\ensuremath{\D^0}}\xspace}
\def\Dzb     {{\ensuremath{\Dbar{}^0}}\xspace}

\def\Dpm     {{\ensuremath{\D^\pm}}\xspace}

\def\Dstar   {{\ensuremath{\D^*}}\xspace}

\def\B       {{\ensuremath{\PB}}\xspace}
\def\Bbar    {{\ensuremath{\kern 0.18em\overline{\kern -0.18em \PB}{}}}\xspace}

\def\BorBbar    {\kern 0.18em\optbar{\kern -0.18em B}{}\xspace}
\def\Bz      {{\ensuremath{\B^0}}\xspace}

\def\Bu      {{\ensuremath{\B^+}}\xspace}
\def\Bub     {{\ensuremath{\B^-}}\xspace}
\def\Bp      {{\ensuremath{\Bu}}\xspace}
\def\Bm      {{\ensuremath{\Bub}}\xspace}
\def\Bpm     {{\ensuremath{\B^\pm}}\xspace}

  \def\Y#1S{\ensuremath{\PUpsilon{(#1S)}}\xspace}

\def\Lbar        {{\ensuremath{\kern 0.1em\overline{\kern -0.1em\PLambda}}}\xspace}
\def\LorLbar    {\kern 0.18em\optbar{\kern -0.18em \PLambda}{}\xspace}

\def\to                 {\ensuremath{\rightarrow}\xspace}

\def\CP                {{\ensuremath{C\!P}}\xspace}

\def\AT#1     {\ensuremath{A_{\mathrm{T}}^{#1}}\xspace}           

\def\C#1      {\ensuremath{\mathcal{C}_{#1}}\xspace}                       
\def\Cp#1     {\ensuremath{\mathcal{C}_{#1}^{'}}\xspace}                    
\def\Ceff#1   {\ensuremath{\mathcal{C}_{#1}^{\mathrm{(eff)}}}\xspace}        
\def\Cpeff#1  {\ensuremath{\mathcal{C}_{#1}^{'\mathrm{(eff)}}}\xspace}       
\def\Ope#1    {\ensuremath{\mathcal{O}_{#1}}\xspace}                       
\def\Opep#1   {\ensuremath{\mathcal{O}_{#1}^{'}}\xspace}                    

\newcommand{\ket}[1]{\ensuremath{|#1\rangle}}              

\newcommand{\tev}{\ifthenelse{\boolean{inbibliography}}{\ensuremath{~T\kern -0.05em eV}}{\ensuremath{\mathrm{\,Te\kern -0.1em V}}}\xspace}
\newcommand{\gev}{\ensuremath{\mathrm{\,Ge\kern -0.1em V}}\xspace}
\newcommand{\mev}{\ensuremath{\mathrm{\,Me\kern -0.1em V}}\xspace}
\newcommand{\kev}{\ensuremath{\mathrm{\,ke\kern -0.1em V}}\xspace}
\newcommand{\ev}{\ensuremath{\mathrm{\,e\kern -0.1em V}}\xspace}
\newcommand{\gevc}{\ensuremath{{\mathrm{\,Ge\kern -0.1em V\!/}c}}\xspace}
\newcommand{\mevc}{\ensuremath{{\mathrm{\,Me\kern -0.1em V\!/}c}}\xspace}
\newcommand{\gevcc}{\ensuremath{{\mathrm{\,Ge\kern -0.1em V\!/}c^2}}\xspace}
\newcommand{\gevgevcccc}{\ensuremath{{\mathrm{\,Ge\kern -0.1em V^2\!/}c^4}}\xspace}
\newcommand{\mevcc}{\ensuremath{{\mathrm{\,Me\kern -0.1em V\!/}c^2}}\xspace}

\def\cm   {\ensuremath{\mathrm{ \,cm}}\xspace}

\def\mm   {\ensuremath{\mathrm{ \,mm}}\xspace}

\def\ci {\ensuremath{\mathrm{ \,Ci}}\xspace}

\def\gsim{{~\raise.15em\hbox{$>$}\kern-.85em
          \lower.35em\hbox{$\sim$}~}\xspace}
\def\lsim{{~\raise.15em\hbox{$<$}\kern-.85em
          \lower.35em\hbox{$\sim$}~}\xspace}

\def\tell1  {TELL1\xspace}
\def\ukl1   {UKL1\xspace}

\usepackage{jheppub} 
\usepackage{scalerel}
\usepackage[T1]{fontenc} 
\usepackage{subcaption}

\usepackage[displaymath, mathlines,running]{lineno}

\newcommand*\patchAmsMathEnvironmentForLineno[1]{%
\expandafter\let\csname old#1\expandafter\endcsname\csname #1\endcsname
\expandafter\let\csname oldend#1\expandafter\endcsname\csname
end#1\endcsname
 \renewenvironment{#1}%
   {\linenomath\csname old#1\endcsname}%
   {\csname oldend#1\endcsname\endlinenomath}%
}
\newcommand*\patchBothAmsMathEnvironmentsForLineno[1]{%
  \patchAmsMathEnvironmentForLineno{#1}%
  \patchAmsMathEnvironmentForLineno{#1*}%
}
\AtBeginDocument{%
\patchBothAmsMathEnvironmentsForLineno{equation}%
\patchBothAmsMathEnvironmentsForLineno{align}%
\patchBothAmsMathEnvironmentsForLineno{flalign}%
\patchBothAmsMathEnvironmentsForLineno{alignat}%
\patchBothAmsMathEnvironmentsForLineno{gather}%
\patchBothAmsMathEnvironmentsForLineno{multline}%
\patchBothAmsMathEnvironmentsForLineno{eqnarray}%
}

\title{\texorpdfstring{\boldmath \CP violation and material interaction of neutral kaons in measurements of the CKM angle $\gamma$ using $\Bpm\to\D\Kpm$ decays where $\D\to\KS\pi^+\pi^-$}
{CP violation and material interaction of neutral kaons in measurements of the CKM angle gamma using B+-->DK+- decays where D->KSpi+pi-}
}

\author{M. Bj\o rn and S. Malde,}

\affiliation{University of Oxford,\\Oxford,\\United Kingdom}

\emailAdd{mikkel.bjoern@physics.ox.ac.uk}
\emailAdd{sneha.malde@physics.ox.ac.uk}

\abstract{

    As measurements of the CKM angle $\gamma$ in decays of $b$-hadrons become increasingly precise, it is important to consider the impact of processes that affect secondary and tertiary decay products and can contribute to the observed \CP violation. The golden decay mode used to measure $\gamma$ is $\B\to\D\kaon$, where $\D\to\KS\pip\pim$. Due to the presence of a \KS meson in the final state, $\gamma$ measurements based on this mode are affected by neutral kaon \CP violation and matter regeneration, and this study examines the potential size of the impact. Previous studies determine the impact of kaon \CP violation to be a few degrees in size, but make simplifying assumptions about how experimental measurements are implemented. These assumptions are lifted in the present study, and the matter regeneration effect is included for the first time. The results are presented within the context of $\gamma$ measurements to be made using the \lhcb and \belle II detectors. It is found that the expected biases due to ignoring the effects of \CP violation and matter regeneration in neutral kaons is small and less than $0.5^\circ$ degrees in all experimental scenarios considered.
 
}

\notoc
\begin{document} 

\maketitle

\flushbottom
\newpage
\setcounter{page}{1}

\section{Introduction} 
\label{sec:introduction}

The \CP-violating phase 
$\gamma \equiv \arg \left({-V_{ud}^{\phantom{\ast}}V_{ub}^\ast}/{V_{cd}^{\phantom{\ast}}V_{cb}^\ast}\right)$
of the Cabibbo-Kobayashi-Maskawa (CKM) matrix~\cite{Cabibbo:1963yz,Kobayashi:1973fv} in the Standard Model (SM) is the only CKM angle that can easily be measured via tree-level processes. It therefore provides an important SM benchmark measurement that can be compared with estimates based on other CKM observables, more likely to be affected by new physics effects~\cite{Blanke:2018cya}. Interference between $\B\to\Dz\kaon$ and $\B\to\Dzb\kaon$ decay amplitudes provide an important way to measure $\gamma$. Measurements have been made in a range of $\B\to\D\kaon$ modes~\cite{Gammacombo2016,Gammacombo2018,HFLAV16,UTfit-UT}, which are combined to provide the best estimate of $\gamma$ from direct measurements, with a current uncertainty of about $5^\circ$~\cite{HFLAV16,UTfit-UT,CKMfitter2015}.

An important contribution comes from $\BtoDK$ decays in which $\D\to\KS\pi^+\pi^-$, where \D describes a superposition of \Dz and \Dzb states. In this decay, $\gamma$ can be determined by analysing the phase-space distribution of the \D decay, as proposed in Refs.~\cite{BONDARGGSZ,BPMODIND1,BPMODIND2} and~\cite{GGSZ}. This approach has been used to measure $\gamma$ by the \belle~\cite{BELLEMODIND,BELLE2004,BELLE2006,BELLE2010}, \babar~\cite{BABAR2005,BABAR2008,BABAR2010}, and \lhcb collaborations~\cite{LHCb-PAPER-2014-017,LHCb-PAPER-2016-007,LHCb-PAPER-2014-041,LHCb-PAPER-2016-006,LHCb-PAPER-2018-017}, with the latest measurement from \lhcb~\cite{LHCb-PAPER-2018-017} constituting the most precise stand-alone $\gamma$ measurement to date. The \lhcb and \belle II collaborations both expect to reach a precision of at least $\sigma(\gamma)\lsim 3^\circ$ in measurements that use the method during the coming decade, with \lhcb expecting a statistical uncertainty below $1^\circ$ in its proposed Upgrade II phase~\cite{TheLHCbCollaboration:2320509,BELLE2-PUB-2018-001}.

However, the presence of a \KS meson in the final state constitutes an additional source of \CP violation in $\Bpm\to\D(\to\KS \pi^+\pi^-)K^\pm$ decays, due to \CP violation in the neutral kaon system. The effect has been broadly ignored experimentally, as the statistical uncertainty has so far been much larger than the expected induced bias on the measured value of $\gamma$. Previous studies have estimated this bias to be up to a few degrees~\cite{YuvalKsCPV}. 
In Ref.~\cite{YuvalKsCPV}, the estimates are based on  phase-space-integrated yield asymmetries, however the current experimental approach is based on \CP asymmetries within the phase-space distribution of signal decays. This paper focuses on the latter approach, and removes some of the assumptions made in Ref.~\cite{YuvalKsCPV}.
Furthermore, material interaction of the neutral kaon in the detector can lead to experimental signatures that mimic those of \CP violation, and thus introduce an additional source of measured \CP asymmetry, which can be of a similar size to the asymmetry from inherent \CP violation in the neutral kaon system~\cite{LHCb-PAPER-2014-013,LHCbKSCPVNote}. Therefore, a detailed understanding of both effects will be necessary as the statistical uncertainties on $\gamma$ measurements based on $\Bpm\to\D(\to\KS \pi^+\pi^-)K^\pm$ decays enter the few degree range.

\section{\texorpdfstring
{Measurements of $\gamma$ using $\Bpm\to\D(\to\KS \pi^+ \pi^-)\Kpm$ decays}
{Model independent measurements of gamma using B->DK, D->Kspipi decays}} 
\label{sec:Theoretical_background}
\label{sub:measuring_gamma_using_ggz}


\renewcommand{\Re}{\text{Re}}
\renewcommand{\Im}{\text{Im}}

\renewcommand{\g}{\ensuremath{\gamma}\xspace}
\newcommand{\rB}{\ensuremath{r_\B}\xspace}
\newcommand{\dB}{\ensuremath{\delta_\B}\xspace}
\renewcommand{\ci}{\ensuremath{c_i}\xspace}
\newcommand{\si}{\ensuremath{s_i}\xspace}
\newcommand{\Ki}{\ensuremath{K_i}\xspace}
\newcommand{\Kmi}{\ensuremath{K_{-i}}\xspace}
\newcommand{\Kpi}{\ensuremath{K_{+i}}\xspace}

\newcommand{\sm}{\ensuremath{s_-}\xspace}
\renewcommand{\sp}{\ensuremath{s_+}\xspace}
\newcommand{\smpLong}{\ensuremath{s_-,s_+}\xspace}
\newcommand{\spmLong}{\ensuremath{s_+,s_-}\xspace}
\newcommand{\smp}{\ensuremath{s_{-+}}\xspace}
\newcommand{\spm}{\ensuremath{s_{+-}}\xspace}
\newcommand{\AB}{\ensuremath{A_\B}\xspace}
\newcommand{\AKS}{\ensuremath{A_{\KS}}\xspace}
\newcommand{\ADS}{\ensuremath{A_{\rm S}^\D}\xspace}
\newcommand{\ADbS}{\ensuremath{A_{\rm S}^\Db}\xspace}
\newcommand{\ADL}{\ensuremath{A_{\rm L}^\D}\xspace}
\newcommand{\ADbL}{\ensuremath{A_{\rm L}^\Db}\xspace}
\newcommand{\ADorDbS}{\ensuremath{A_{\rm S}^{\scaleto{\DorDbar}{9pt}}}\xspace}
\newcommand{\ADorDbL}{\ensuremath{A_{\rm L}^{\scaleto{\DorDbar}{9pt}}}\xspace}
\newcommand{\ADorDbonetwo}{\ensuremath{A_{1/2}^{\scaleto{\DorDbar}{9pt}}}\xspace}
\newcommand{\ADorDbSL}{\ensuremath{A_{\rm S(L)}^{\scaleto{\DorDbar}{9pt}}}\xspace}
\newcommand{\cASm}{\ensuremath{\mathcal{A}_{\rm S}^-}\xspace}
\newcommand{\cASp}{\ensuremath{\mathcal{A}_{\rm S}^+}\xspace}
\newcommand{\cALm}{\ensuremath{\mathcal{A}_{\rm L}^-}\xspace}
\newcommand{\cALp}{\ensuremath{\mathcal{A}_{\rm L}^+}\xspace}
\newcommand{\Kone}{\ensuremath{K_\rm{1}}\xspace}
\newcommand{\Ktwo}{\ensuremath{K_\rm{2}}\xspace}

This section introduces the theory of $\gamma$ measurements based on $\Bpm\to\D(\to\KS \pi^+ \pi^-)\Kpm$ decays and the current experimental practice in Section~\ref{sub:current_practice}, then describes the effects of neutral kaon \CP violation and material interaction in Section~\ref{sub:cp_violation_and_material_interaction_of_neutral_kaons}, and gives an estimate of the expected impact on experimental results in Section~\ref{sub:yields_to_lowest_order}.

\subsection{Theory and current experimental practice} 
\label{sub:current_practice}

The amplitude for the decay $\Bm\to\D(\to\KS(\to \pip\pim) \pi^+ \pi^-)\Km$, denoted $\cASm$, can be written as a superposition of the amplitude for a $\Bm\to\Dz\Km$ decay and the suppressed $\Bm\to\Dzb\Km$ decay
\begin{align} \label{eq:A_Bminus}
    \cASm(s_-, s_+)
    &= \AB \AKS \left(\ADS(\smpLong) + \rB \exp[i(\dB - \g)]\ADbS(\smpLong)\right),
\end{align}
where \rB is the ratio of the magnitudes of the $\Bm\to\Dzb\Km$ and $\Bm\to\Dz\Km$ amplitudes, \dB is the strong phase between them, \AB is the amplitude of the $\Bm\to\Dz\Km$ decay, \AKS is the amplitude of the $\KS\to\pip\pim$ decay, \sm and \sp are the squared invariant masses of the $\KS\pim$ and $\KS\pip$ particle combinations, respectively, and the \D decay amplitudes are defined as
\begin{align}\label{eq:ADDb_definition}
\ADorDbSL (\smpLong)= A(\DorDbar^0\to K^0_\text{S(L)} \pi^+\pi^-).
\end{align}
Biases from the effects of $\Dz-\Dzb$ mixing are potentially significant, but can be confined to 0.1$^\circ$ with an appropriate measurement strategy. These effects are described in Ref.~\cite{Dmixing} and are not discussed further in this study. Direct \CP violation in the \D decay is assumed to be negligible, as the effect is expected to be very small for $\D\to\KS\pip\pim$ decays in the SM, and has been analysed in Ref.~\cite{Bondar2013}.
Under the further assumption that \KS is a \CP eigenstate, the \D decay amplitudes satisfy 
\begin{align}\label{eq:KS_symmetry}
     \ADbS(\smp)=\ADS(\spm), 
 \end{align} where the short-hand notation $(s_{-+})=(s_-,s_+)$ and $(s_{+-})=(s_+,s_-)$ is employed to simplify equations. The differential $\Bm\to\D(\to\KS \pi^+ \pi^-)\Km$ decay rate to a given point in the \D decay phase space is
\begin{align} \label{eq:Gamma_Bminus}
    \rm {d} \Gamma^-(\smp) &\propto |\cASm|^2 = |\AB|^2|\AKS|^2 \left[|\ADS(\smp)|^2 + \rB^2 |\ADS(\spm)|^2 + 2\rB |\ADS(\smp)||\ADS(\spm)|\right .
    \notag \\
    &\qquad \left. \times \left(\cos[-\Delta\delta_D(\smp)]\cos[\dB-\g]-\sin[-\Delta\delta_D(\smp)]\sin[\dB-\g]\right)\right].
\end{align}
Here, $\Delta\delta_\D(\smp) = \delta_D(\smp) - \delta_D(\spm)$, where $\delta_D(\smp)$ denotes the phase of $\ADS(\smp)$. The equivalent expressions to Eq.~\eqref{eq:A_Bminus} and Eq.~\eqref{eq:Gamma_Bminus} for $\Bp\to\D(\to\KS \pi^+ \pi^-)\Kp$ decays are obtained via the substitutions $\g\to-\g$ and $\ADS(\smp)\leftrightarrow\ADbS(\smp)$, where the latter substitution is equivalent to $\ADS(\smp)\leftrightarrow\ADS(\spm)$.

\begin{figure}[tb]
    \centering
    \includegraphics{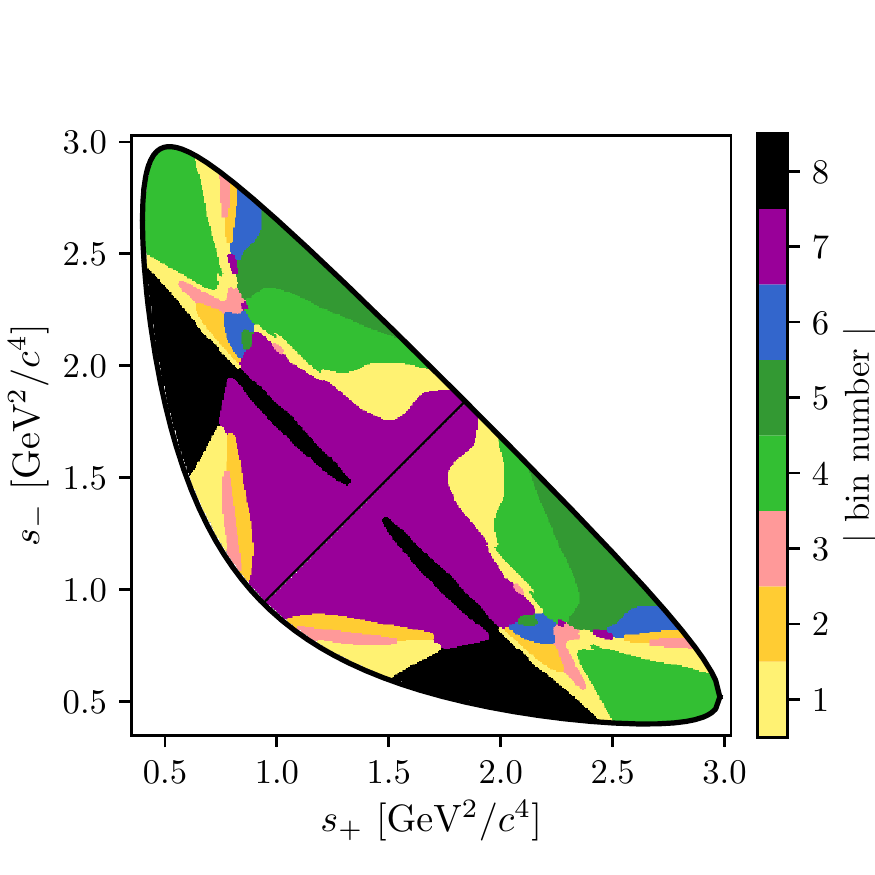}
    \caption{The \emph{optimal binning scheme} of the $\D\to\KS\pip\pim$ phase space~\cite{CLEOCISI}.}
    \label{fig:optimal_binning_scheme}
\end{figure}

Based on Eq.~\eqref{eq:Gamma_Bminus}, it is possible to measure \g (and the nuisance parameters \rB and \dB) from the phase-space distribution of $\Bpm\to\D(\to\KS \pi^+ \pi^-)\Kpm$ decays, given knowledge of $\delta_D(\smp)$ or $\Delta\delta_D(\smp)$. A series of \g measurements have used amplitude models of the \D decay to describe $\Delta\delta_D(\smp)$ \cite{BABAR2005,BABAR2008,BABAR2010, BELLE2004,BELLE2006,BELLE2010,LHCb-PAPER-2014-017,LHCb-PAPER-2016-007}. More recently, a model-independent approach, which is critical when looking for evidence of beyond-the-SM effects~\cite{BELLEMODIND,LHCb-PAPER-2014-041,LHCb-PAPER-2016-006,LHCb-PAPER-2018-017}, has been favoured. The \D decay phase space is split into $2\times N$ regions, or bins, numbered $i=-N$ to $N$ (omitting zero) that are defined to be symmetric around $\sm=\sp$, with $i>0$ for $s_- >s_+$. The binning scheme used in current experimental measurements of $\Bpm\to\D\Kpm$ decays~\cite{BELLEMODIND,LHCb-PAPER-2014-041,LHCb-PAPER-2018-017} has $N=8$ and is shown in Fig.~\ref{fig:optimal_binning_scheme}. Then, the average of $\cos \Delta\delta_D(\smp)$ over bin $i$ of the \D decay phase space is
\begin{align}\label{eq:ci_si}
    \ci = \frac
    {\int_i \text{d}s^2 |\ADS(\smp)||\ADS(\spm)|\cos[\Delta\delta_D(\smp)]}
    {\sqrt{\int_i \text{d}s^2 |\ADS(\smp)|^2}\sqrt{\int_i \text{d}s^2 |\ADS(\spm)|^2}},
\end{align}
where $\int_i\text{d}s^2$ denotes integration over bin $i$ of the \D decay phase space, and with an analogous definition of \si, the average of $\sin \Delta\delta_D(\smp)$.  Integrating Eq.~\eqref{eq:Gamma_Bminus}, and the corresponding expression for $\Bp\to\D\Kp$ decays, the yields of $\Bpm\to\D(\to\KS \pi^+\pi^-)\Kpm$ in bin $i$ are
\begin{align}
\begin{split}    \label{eq:base_yields}
    N^-_i &= h_B^-\left(\Ki + \rB^2\Kmi + 2\sqrt{\Ki\Kmi}\left(\ci\xm+\si\ym\right)\right), \\
    N^+_i &= h_B^+\left(\Kmi + \rB^2\Ki + 2\sqrt{\Ki\Kmi}\left(\ci\xp-\si\yp\right)\right),
\end{split}
\end{align}
in terms of the integrals 
\begin{align}\label{eq:base_ki}
    K_i &= \frac{1}{N_K}\int_i\text{d}s^2 |\ADS(\smp)|^2, &
    N_K &=\int\text{d}s^2 |\ADS(\smp)|^2,
\end{align}the normalisation constants $h^\pm_B$, and the \CP violation observables ${\xpm = \rB \cos (\dB \pm \g)}$ and $ \ypm = \rB \sin (\dB \pm \g)$.

In model-independent experimental measurements of \g, the signal decay yields in each phase-space bin are expressed via Eq.~\eqref{eq:base_yields} and \xy are determined via a maximum likelihood fit to the data. As Eq.~\eqref{eq:base_yields} leads to 32 observable yields but 36 unknown parameters, it is necessary to use some external information. It is possible to measure \ci and \si using quantum correlated \Dz\Dzb pairs produced at the $\psi(3770)$ resonance \cite{BPMODIND1,BPMODIND2}. Such measurements have been made by the \cleo collaboration~\cite{CLEOCISI}, and these have been employed in a range of \g measurements \cite{BELLEMODIND,LHCb-PAPER-2014-041,LHCb-PAPER-2016-006,LHCb-PAPER-2018-017}. More precise measurements of \ci and \si are expected from the BESIII collaboration, and further measurements could also be made from an analysis of charm mixing \cite{CTGWcsFromD}, or from reconstructed decays of the $\psi(3770)$ meson at LHCb~\cite{VanyaPsi3770}. The \Ki parameters can also be measured in external samples. It is advantageous to determine the \Ki parameters from a sample that has as similar kinematics to the $\Bpm\to\D\Kpm$ sample as possible, in order to automatically include corrections from experimental effects, such as reconstruction efficiency and resolution. A sample of flavour-tagged \D decays is commonly used. Given the current yields of $B$ decays, both the strong phase and \Ki parameters are taken from external input in order to maximise the sensitivity to \xy.

The interference in $\Bpm\to\D(\to\KS\pi^+ \pi^-)\Kpm$ decays presents itself primarily in different distributions over the \D-decay phase space between signal decays originating from \Bp and \Bm mesons. From an experimental point of view this is highly desirable, as production and detection asymmetries that affect the phase-space-integrated yields can be ignored. In addition to the asymmetry in the distribution over the phase space there is a further \CP asymmetry in the phase-space-integrated yields, expected to be around 1\,\%. This \CP asymmetry is very challenging to measure with useful precision, due to the limited sample sizes currently available and possible biases of a similar magnitude from production and detection asymmetries. Hence, in most studies it is ignored and makes no contribution to the overall determination of \g.

\subsection{\CP violation and material interaction of neutral kaons} 
\label{sub:cp_violation_and_material_interaction_of_neutral_kaons}

The derivation in the preceding section assumed that \KS is a \CP eigenstate, however this is known to be an approximation. Including \CP violation in the neutral kaons, the mass eigenstates \KS and \KL are given by
\begin{align}
    \ket{\KS}   &= \frac{1}{\sqrt{ 1+|\epsilon|^2}} \left[\ket{K_1} + \epsilon \ket{K_2}\right], &
    \ket{\KL}   &= \frac{1}{\sqrt{ 1+|\epsilon|^2}} \left[\ket{K_2} + \epsilon \ket{K_1}\right],
\end{align}
in terms of the \CP even (odd) eigenstates $K_{1(2)}=(\Kz+(-)\Kzb)/\sqrt{2}$ and the \CP violation parameter
$
\epsilon = A(\KL\to\pip\pim)/A(\KS\to\pip\pim)
$, measured to be~\cite{PDG2018}  
\begin{align}\label{eq:PDG_epsilon}
    |\epsilon|=(2.228\pm 0.011)\times 10^{-3}, \qquad \arg \epsilon = (43.52\pm 0.05)^\circ.
\end{align}
The phase-convention $\hat C\hat P \ket{\Kz}=\ket{\Kzb}$ is used and direct \CP violation in the kaon decay is ignored, as the effect is three orders of magnitude smaller than indirect \CP violation~\cite{PDG2018}. The non-zero amplitude for the $\KL\to\pip\pim$ decay introduces a direct dependence on the $\ADorDbL$ amplitudes.  Whereas Eq.~\eqref{eq:Gamma_Bminus} assumed that $\text{d} \Gamma^-(t, \smp) \propto \left|\psi^-_\text{S}(t, \smp)\right|^2$, the actual decay rate satisfies
\begin{align}\label{eq:Gamma_Bminus_with_KL}
    \text{d} \Gamma^-(t, \smp) &\propto \left|\psi^-_\text{S}(t, \smp) + \epsilon \psi^-_\text{L}(t, \smp)\right|^2,
\end{align}
where $\psi^-_\text{S}$ and $\psi^-_\text{L}$ are the \KS and \KL components of the neutral kaon state.\footnote{The time dependence of Eq.~\eqref{eq:Gamma_Bminus_with_KL} is kept implicit in Eq.~\eqref{eq:Gamma_Bminus}, as it contributes a factor that is constant over phase-space in the $\epsilon=0$ approximation.} The $\ADorDbL$ amplitudes do not satisfy Eq.~\eqref{eq:KS_symmetry}, but instead $\ADL(\smp)\simeq-\ADbL(\spm)$, and therefore the presence of the \KL term leads to corrections to the yield expressions in Eq.~\eqref{eq:base_yields}. In an experimental setting, the dependence on the $\ADorDbL$ amplitudes is further enhanced by material interactions of the neutral kaon, because different nuclear interaction strengths of the \Kz and \Kzb mesons introduce a non-zero $\KS\leftrightarrow\KL$ transition amplitude for neutral kaons traversing a detector segment. This effect was predicted early in the history of kaon physics~\cite{PaisPiccioni1955} and is commonly denoted \emph{kaon regeneration}. The general expression for the time dependent neutral kaon state components is~\cite{Good1957,Fetscher1996}
\begin{align} 
\begin{split} \label{eq:mat_time_dep}
\psi_{\rm{S}}(t, \smp) &= e^{-i\Sigma t}\left( \psi^0_{\rm S}(\smp)\cos \Omega t  
                + \frac{i}{2\Omega}\left(\Delta \lambda \psi^0_{\rm S}(\smp) - \Delta \chi \psi^0_{\rm L}(\smp)\right) \sin \Omega t \right) ,\\
\psi_{\rm{L}}(t, \smp) &= e^{-i\Sigma t}\left( \psi^0_{\rm L}(\smp)\cos \Omega t  
                - \frac{i}{2\Omega}\left(\Delta \lambda \psi^0_{\rm L}(\smp) + \Delta \chi \psi^0_{\rm S}(\smp)\right) \sin \Omega t \right) ,
\end{split}
\end{align} in terms of the parameters
\begin{align} 
\begin{split}
 \label{eq:mat_param}
\Delta \chi &= \chi - \bar \chi, \\
\Delta \lambda &= \lambda_{\rm L} - \lambda_{\rm S} = (m_{\rm L} - m_{\rm S}) - \frac{i}{2}(\Gamma_{\rm L} - \Gamma_{\rm S}),
\\ 
 \Sigma & = \frac{1}{2}\left(\lambda_{\rm S} + \lambda_{\rm L} + \chi + \bar \chi\right), \\
 \Omega &= \frac{1}{2}\sqrt{\Delta \lambda^2 + \Delta \chi^2},
\end{split}
\end{align}
where $m_\text{S(L)}$ and $\Gamma_\text{S(L)}$ are the mass and decay width of the \KS (\KL) mass eigenstates, and the parameters $\chi$ and $\bar\chi$ describe the material interaction of the \Kz and \Kzb flavour eigenstates. The $\chi$ $(\bar \chi)$ parameter is proportional to the forward scattering amplitude of a \Kz (\Kzb) meson in a traversed material. In Eq.~\eqref{eq:mat_time_dep}, $\psi^0_\text S$ and $\psi^0_\text L$ are the initial ${\KS} $ and ${\KL}$ components of the neutral kaon state, which depend on the phase-space coordinates of the \D decay: $\psi^0_\text {S/L} \propto \mathcal{A}_\text{S/L}(\smp)$. Thus, for $\Delta\chi\neq 0$, $\psi_\text{S}(t)$ depends on $\mathcal{A}_\text{L}(\spm)$ irrespective of the $\KL\to\pi^+\pi^-$ decay, due to kaon regeneration.

In addition, the relations $\ADbS(\smp)=\ADS(\spm)$ and $\ADbL(\smp)=-\ADL(\spm)$ are not exact for $\epsilon \neq 0$, as \KS and \KL  are not exact \CP eigenstates. This leads to further corrections to the yield expressions in Eq.~\eqref{eq:base_yields}. It it beneficial to express $A^\D_\text{S(L)}$ in terms of the amplitudes $A^\D_{1(2)}$, defined analogously to  Eq.~\eqref{eq:ADDb_definition} but for the \CP even (odd) eigenstates $K_1$ ($K_2$). After the decay of a \Dz meson to a neutral kaon, the kaon state is
\begin{align}
\begin{split}
    \psi^0 &= A^\D_1 \ket {K_1} + A^\D_2\ket{K_2} \\
    &= N\left[(A^\D_1-\epsilon A^\D_2)\ket\KS
    + (A^\D_2-\epsilon A^\D_1)\ket\KL\right],
\end{split}
\end{align}
with the normalisation constant $N=\sqrt{1+|\epsilon|^2}/(1-\epsilon^2)$. Thus it can be seen that
\begin{align}
\begin{split}\label{eq:A12toKS}
    A^\D_\text{S}(\spm) &= N\left[(A^\D_1(\spm)-\epsilon A^\D_2(\spm))\right], \\
    A^\D_\text{L}(\spm) &= N\left[(A^\D_2(\spm)-\epsilon A^\D_1(\spm))\right],
\end{split}
\end{align}
with an analogous expression for the \Dzb decay amplitudes. Therefore, the  generalised relations between the \Dz and \Dzb amplitudes are
\begin{align}
\begin{split}\label{eq:DDbar_relations}
    \ADbS(\spm) &= \phantom{-}N[A^\Dbar_1(\spm)-\epsilon A^\Dbar_2(\spm)]
    \\
    &= \phantom{-}N[A^\D_1(\smp)+\epsilon A^\D_2(\smp)]  = \phantom{-}\ADS(\smp) + 2N\epsilon A^\D_2(\smp),
    \\
    \ADbL(\spm) &= \phantom{-}N [A^\Dbar_2(\spm)-\epsilon A^\Dbar_1(\spm) ]
    \\
    & = -N[A^\D_2(\smp)+\epsilon A^\D_1(\smp)] = -\ADL(\smp) - 2N\epsilon A^\D_1(\smp).
\end{split}
\end{align} 

In order to calculate the full corrections to the yield expressions in Eq.~\eqref{eq:base_yields}, models of $A_\text{S}^D$ and $A_\text{L}^D$ (or $A_\text{1}^D$ and $A_\text{2}^D$) are needed. While there are several amplitude models available to describe the decay amplitude $A(\Dz\to\KS\pi^+\pi^-)$~\cite{BABAR2005,BABAR2008,BABAR2010,BELLE2010,DalitzModel18}, no models have been published for $\Dz\to\KL \pi^+\pi^-$  decays. However, following the assumptions laid out in Ref.~\cite{CLEOCISI}, the amplitudes $A^\D_1(\spm)$ and $A^\D_2(\spm)$ can be related, allowing both qualitative and quantitative estimates of the bias effects to be made with existing $\Dz\to\KS\pip\pim$ models. In the isobar formalism, the decay amplitude $A(\Dz\to K_1 \pi^+\pi^-)$ is expressed as a non-resonant constant amplitude plus a sum of resonances
\begin{align}\label{eq:amplitude}
    A(\Dz\to K_1 \pi^+\pi^-) = k_{NR} + \sum_{CF} k_i R^i(s_{K\pi^-}) + \sum_{DCS} k_j R^j(s_{K\pi^+}) + \sum_{R_{\pi\pi}} k_k R^k(s_{\pi^+\pi-}).
\end{align}
The resonances are split into Cabibbo-favoured (CF) \Kstarm resonances, doubly Cabibbo-suppressed (DCS) \Kstarp resonances and $\pi\pi$ resonances. The $R$ functions are taken to describe all kinematical dependence and are well described in eg. Refs.~\cite{BABAR2005,DalitzModel18} and references therein. In modern models, the $\pi\pi$ and $K\pi$ $S$-wave components are modelled via the $K$-matrix formalism and LASS parametrisations, respectively, instead of sums of individual resonances~\cite{DalitzModel18}. This does not alter the arguments below, as the $R$ functions of Eq.~\eqref{eq:amplitude} can equally well represent such terms. The CF resonances couple to the \Kzb component of $K_1(\propto \Kz+\Kzb)$, and therefore the corresponding $k_i$ in the $K_2(\propto \Kz-\Kzb)$ amplitude will have a relative minus sign. The DCS resonances couple to the \Kz component of $K_1$, and so the corresponding $k_j$ in the $K_2$ amplitude will have a relative plus sign. For the $h^+h^-$ resonances, there will be a coupling to both the \Kz and \Kzb components, however the coupling to the \Kz component is expected to be suppressed with a Cabibbo suppression factor $r_ke^{i\delta_k}$, where $r_k\simeq\tan^2\theta_C \simeq 0.05$ is determined by the Cabibbo angle $\theta_C$ and $\delta_k$ can take any value.  Therefore, the $k_k$ for these resonances have a relative $-(1-2r_ke^{i\delta_k})$ factor in the $K_2$ amplitude. The same effect leads to the differences in decay rates between $\Dz\to\KS\piz$ and $\Dz\to\KL\piz$ decays~\cite{TanThetaCTheory,TanThetaCCLEO}. An important consequence of these substitution rules is that
\begin{align}\label{eq:A1_A2_relation}
    A^\D_2(\spm) = -A^\D_1(\spm)+r_A \Delta A(\spm),
\end{align}
where $r_A\simeq\tan^2\theta_C $ and $\Delta A(\spm)\sim A^\D_1(\spm)$ are of the same order of magnitude (at least when averaged over the bins used in $\gamma$ measurements). 
This relation is sufficient to make the qualitative arguments of Section~\ref{sub:yields_to_lowest_order}, while the full set of substitution rules above are used in the quantitative studies of Section~\ref{sec:numerical_studies}.

{}
\subsection{
\texorpdfstring{Impact on $\gamma$ measurements}
{Impact on gamma measurements}} 
\label{sub:yields_to_lowest_order}

With suitable models to calculate \ADorDbSL (or \ADorDbonetwo) and knowledge of $\Delta\chi$ for the materials relevant to an experimental setting, Eqs.~\eqref{eq:Gamma_Bminus_with_KL}, \eqref{eq:mat_time_dep}, and \eqref{eq:DDbar_relations} can be integrated to calculate the expected phase-space bin yields, $N^\pm_i$, including the effects of kaon \CP violation and material interaction. Preliminary to doing this in Section~\ref{sec:numerical_studies}, it is useful to look at the lowest order corrections to Eq.~\eqref{eq:base_yields} in $\epsilon$ and $r_\chi=\frac{1}{2}\frac{\Delta \chi}{\Delta \lambda}$, the dimensionless parameter governing material interactions. For \lhcb and \belle II the average $|r_\chi|\simeq10^{-3}$, as detailed in the Section~\ref{sec:numerical_studies}. The studies of this section are made with the assumption of a flat phase-space efficiency and uniform acceptance over all decay times. Time-acceptance effects will be treated in Section~\ref{sec:numerical_studies}. To first order in $r_\chi$, the expression in Eq.~\eqref{eq:mat_time_dep} simplifies to~\cite{Fetscher1996}
\begin{align}
\begin{split}
        \psi_\text{S}(t, \spm) &= e^{-\frac{i}{2}(\chi + \bar \chi)t}e^{-i \lambda_{\rm S}t}\left( \psi^0_{\rm S}(\spm)
        -r_\chi\left(1-e^{-i\Delta\lambda t} \right)\psi^0_{\rm L}(\spm)\right), \\
    \psi_\text{L}(t, \spm) &= e^{-\frac{i}{2}(\chi + \bar \chi)t}e^{-i \lambda_{\rm L}t}\left( \psi^0_{\rm L}(\spm)  
        +r_\chi\left(1-e^{+i\Delta\lambda t}\right)\psi^0_{\rm S}(\spm)\right).
\end{split}
\end{align}
 In model-independent measurements, the $K_i$ are obtained in a data-driven way using flavour-tagged \D samples, and by averaging over \Dz and \Dzb decays. Therefore it proves beneficial to introduce the parameters
\begin{align}\label{eq:hat_Ki}
\hat K_i &= \frac{1}{1+|\epsilon+r_\chi|^2\frac{\Gamma_\text{S}}{\Gamma_\text{L}}}\left(K_i^{(1)}+|\epsilon+r_\chi|^2\frac{\Gamma_\text{S}}{\Gamma_\text{L}}K_i^{(2)}\right),
\end{align}
in which the $K^{(1/2)}_i$ parameters are phase-space integrals, defined as in Eq.~\eqref{eq:base_ki} but for $A^\D_{1/2}$. The $\hat \Ki$ correspond to the expected measured value of $K_i^\text{meas}=({N^\D_i + N^\Dbar_{-i}})/({\sum_j N^\D_j +N^\Dbar_{-j}})$ to lowest order in $\epsilon$ and $r_\chi$, where $N^\D_i$ $(N^\Dbar_i)$ is the expected yield of flavour tagged \Dz (\Dzb) mesons into bin $i$ of the \D decay phase-space. In fits of amplitude models where both flavour tagged \Dz and \Dzb decays are used to fit the $\D\to\KS\pip\pim$ amplitude, related via Eq.~\eqref{eq:KS_symmetry}, one will effectively fit an amplitude describing $N^\D_i+N^\Dbar_{-i}$, and therefore the arguments below, based on $\hat K_i$, will also hold for model-dependent measurements. Employing Eq.~\eqref{eq:A1_A2_relation} in Eq.~\eqref{eq:hat_Ki}, the expected yields can be written
\begin{align}
\begin{split}\label{eq:final_yields}
    N^-_i&= h_B^{-'} \left( \hat K_{+i} + r_B^2\hat K_{-i} + 2 \sqrt{\hat K_{+i} \hat K_{-i}}(x_- \hat c_i + y_- \hat s_i)   + O(r\epsilon) \right),
    \\
    N^+_i&= h_B^{+'} \left(\hat K_{-i} + r_B^2\hat K_{+i} + 2 \sqrt{\hat K_{+i} \hat K_{-i}}(x_+ \hat c_i - y_+ \hat s_i)   + O(r\epsilon) \right),
        \end{split}
\end{align}
where $O(r\epsilon)$ denotes terms of $O(r_{A}\epsilon)$, $O(r_{B}\epsilon)$, $ O(r_{A}r_\chi)$,  and $ O(r_{B}r_\chi)$. Since $r_B \sim r_A \sim 10^{-1}$ (in $\Bpm\to\D\Kpm$ decays) and $r_\chi\sim\epsilon \sim 10^{-3}$, these terms are all of the same order of magnitude. The new normalisation constants $h_B^{\pm'}=h_B^{\pm}(1+|\epsilon+r_\chi|^2\frac{\Gamma_\text{S}}{\Gamma_\text{L}}\mp\Delta h)$ are defined in terms of
\begin{align}
        \Delta h &= 2\Re[\epsilon+r_\chi] 
    -4\frac{\Gamma_\text{S}}{\Gamma_\text{L}+\Gamma_\text{S}}\frac{\Re[\epsilon+r_\chi] + \mu \Im[\epsilon+r_\chi] }
    {1+\mu^2}, &
    \mu = 2 \frac{m_\text{L}-m_\text{S}}{\Gamma_\text{L}+\Gamma_\text{S}}.
\end{align}
The parameters $(\hat c_i, \hat s_i)$ have been introduced to denote the \emph{measured} average strong-phases, which are expected to differ from $(\ci,\si)$ at $O(\epsilon)$, since neutral kaon \CP violation is not taken into account in the measurements by \cleo. The corrections are thus in the neglected $O(r_B\epsilon)$ terms.
 
Two observations can be made from the expression in~\eqref{eq:final_yields}. The first is that the phase-space \emph{distribution} is only changed at $O(r\epsilon)$ compared to the expression in Eq.~\eqref{eq:base_yields}, if the measured $\hat K_i$ are used in the experimental analysis. As the $\Dz-\Dzb$ interference term that provides sensitivity to $\gamma$ enters at order $O(r_B)$, the impact on $\gamma$ measurements can be expected to be $\Delta\gamma/\gamma\sim O(r\epsilon/r_B)$. For $\B\to\D K$ analyses, where $r_B\simeq0.1$, this is at the permille level, so the induced $\Delta\gamma$ bias can be expected to be smaller than $1^\circ$. This holds true, unless the integrated material interaction, and thereby effective $r_\chi$, varies significantly across the \D-decay phase-space due to experimental effects. However, this is unlikely to be the case in practice, since no significant correlation between the phase-space coordinates and the travel direction of the kaon is expected. 

\begin{figure}[tb]
    \centering{}
    \includegraphics{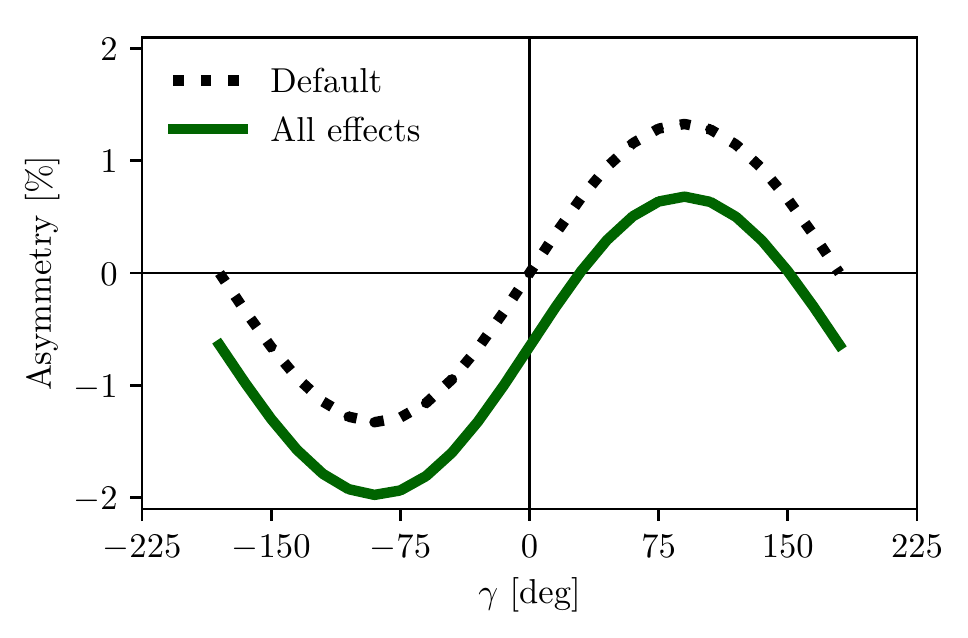}
    \caption{The asymmetry $A_\text{total}$ as a function of $\gamma$ calculated to $O(\epsilon)$ using Eq.~\eqref{eq:global_asym}. The calculation is made using for (black dotted line) the default case where $\Delta h = 0$ and (green) including neutral kaon \CP-violation and material interaction with $r_\chi=\epsilon$.}
    \label{fig:global_asym}
\end{figure}

The second observation relates to potential future measurements of $\gamma$, which may also include sensitivity from the total, phase-space-integrated yield asymmetry
\begin{align}\label{eq:global_asym}
    A_\text{total}=\frac{N^--N^+}{N^-+N^+} = 
    \frac{ 2\sum_{i} c_i\sqrt{\hat K_i\hat K_{-i}} r_B \sin \delta_B \sin \gamma +\Delta h}
    {1 + r_B^2+ 2\sum_ic_i\sqrt{\hat K_i\hat K_{-i}} r_B \cos \delta_B \cos \gamma} + O(r\epsilon),
\end{align}
which was considered in Ref.~\cite{YuvalKsCPV}. In the limit $r_B\to 0$ the expression agrees with the result for the analogous asymmetry in $\Dpm\to\pipm\KS$ decays in Ref.~\cite{Grossman2012}, evaluated to $O(\epsilon)$ for an infinite and uniform time-acceptance. The asymmetry due to \CP violation in the neutral kaon sector, governed by $\Delta h$, is of approximately the same order of magnitude as the asymmetry due to $\gamma$ being non-zero. This is illustrated in Fig.~\ref{fig:global_asym}, where the expression in Eq.~\eqref{eq:global_asym} is plotted in the default case where $\Delta h=0$, using the model in Ref.~\cite{DalitzModel18} to calculate \Ki and \ci, as well as including neutral kaon \CP violation and material interaction effects, calculated using $r_\chi=\epsilon$, with $\epsilon$ taking the value in Eq.~\eqref{eq:PDG_epsilon}. The asymmetry changes significantly when including the latter effects. Therefore, measurements based only on the global asymmetry will suffer relative biases of tens of degrees, not a few degrees, if neutral kaon \CP violation and material interaction is not taken into account. The contribution to $A_\text{total}$ due to \CP violation in the \B decay is an order of magnitude smaller than the $O(r_B)$ expectation described in Ref.~\cite{YuvalKsCPV} because $\sum_ic_i\sqrt{K_iK_{-i}}\simeq 0.1 \ll 1$. The reason is that $\KS\pip\pim$ is not a \CP eigenstate and the strong-phase $\Delta\delta_D(\smp)$ has a non-trivial phase-space dependence. This results in the CF and DCS interference term, which governs the \CP asymmetry, changing sign over phase-space and therefore giving a small contribution to the phase-space-integrated yields.

\section{Expected biases in \lhcb and \belle II} 
\label{sec:numerical_studies}

In order to estimate the effects of \CP violation and material interaction on $\gamma$ measurements that are based on the phase-space distribution of signal decays, the equations of Section~\ref{sub:cp_violation_and_material_interaction_of_neutral_kaons} need to be evaluated to at least $O(r\epsilon)$. Therefore a set of numerical studies are carried out, in which calculations are made to all orders in $\epsilon$, $r_\chi$, $r_A$ and $r_B$. Furthermore, the bias effects depend on the specific detector material budget and time acceptance of a given experiment. The bias is calculated considering the conditions at the two main flavour physics experiments where measurements of \g will be performed in the next decade: \lhcb and \belle II.

\subsection{Time acceptance, momentum distribution, and material parameters} 
\label{sub:material_parameters}
Experiment specific biases are obtained for \lhcb and \belle II, by assuming time acceptances, momentum distributions, and detector geometries typical of the experiments.
The \lhcb experiment is a forward arm spectrometer where the $B$ mesons are produced in proton-proton collisions at 13 TeV. Subsequent decays of the \KS are highly boosted and can occur within different detector subsystems, which leads to two distinct categories of candidates, with different mean lifetimes and material traversed. Therefore  two scenarios are considered for \lhcb:  one in which the decay products of the \KS leave reconstructed tracks in both the silicon vertex detector and downstream tracking detectors (denoted \emph{long-long} or LL), and one in which the decay products of the \KS only leave tracks in the downstream tracking detectors (denoted \emph{down-down} or DD). At \belle II, \B mesons are produced from decays of $\Upsilon(4S)$ mesons, produced in asymmetric \text{electron-positron} collisions. This leads to substantially different decay kinematics in comparison to those found at \lhcb. A single scenario is considered for \belle II, because nearly all the \KS mesons produced in signal decays in Belle II decay within the tracking volume, with more than 90\,\% decaying in the vertex detector according to the studies described below. Thus, three scenarios are considered in total: LL \lhcb, DD \lhcb, and \belle II.

In order to model the experimental time acceptance, the time-dependent integral in Eq.~\eqref{eq:Gamma_Bminus_with_KL} is only carried out over a finite time interval $(\tau_1, \tau_2)$. The intervals are defined for each of the three experimental categories, by requiring that a neutral kaon, if produced at $x=y=z=0$ with momentum $p=(p_T, p_z)$, decays within the relevant part of the corresponding detector. The time acceptance has a significant impact for the \lhcb categories, where some 20\,\% of the kaons escape the tracking stations completely before decaying, whereas the resulting cut-off, $\tau_2$, is large enough in \belle II to have negligible significance. A discussion on the exact requirements placed, and corresponding decay lengths, is found in appendix~\ref{app:detector_details}.

\begin{figure}[tb]
    \centering
    \includegraphics{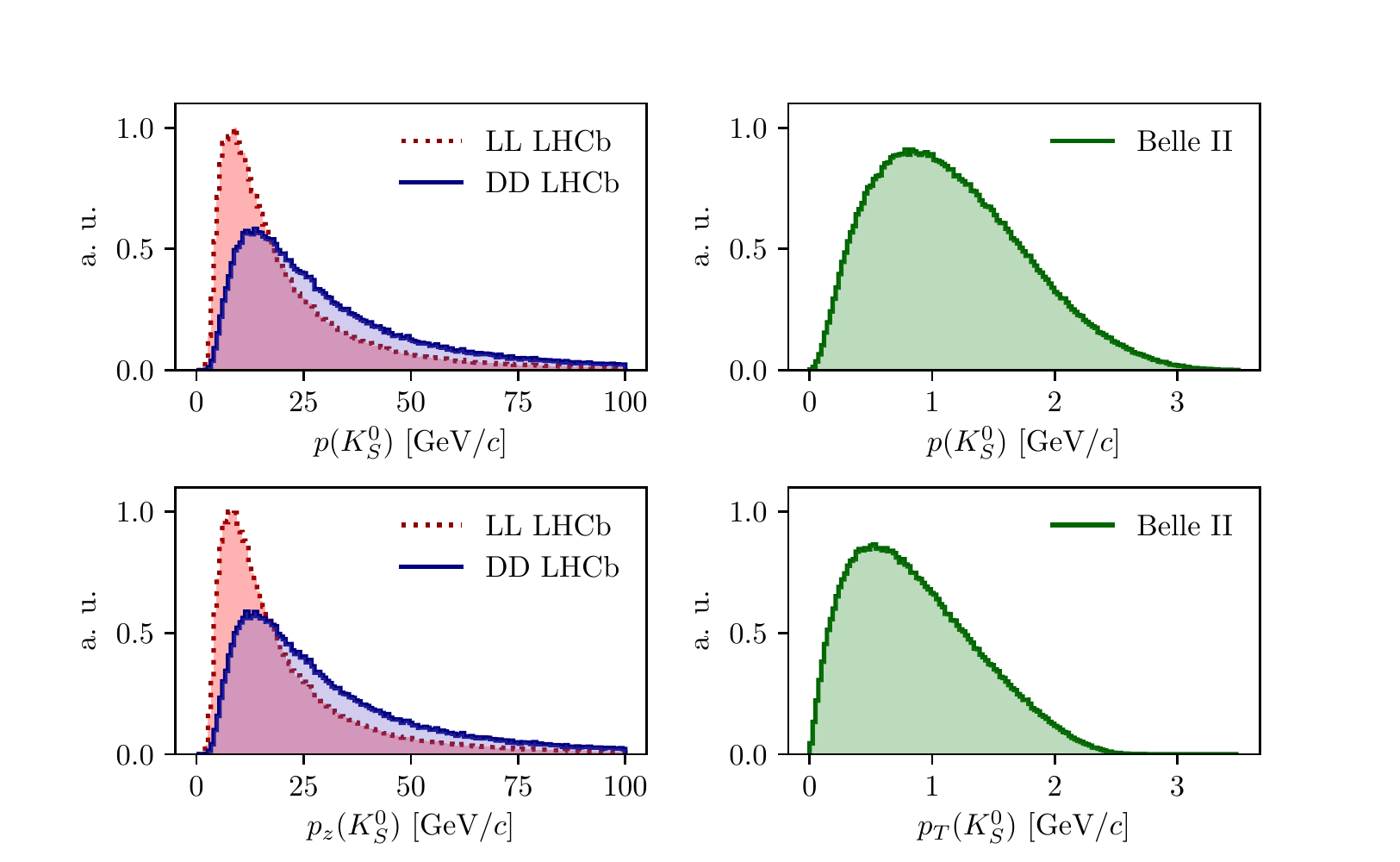}
    \caption{Momentum distributions for the \lhcb (red dotted line) LL and (blue) DD categories, as well as (green) \belle II, obtained using \texttt{RapidSim}.}
    \label{fig:momentum}
\end{figure}

The neutral kaon momentum distribution in \lhcb is obtained using \texttt{RapidSim}~\cite{RapidSim}, which can generate decays of \B mesons with the kinematic distribution found in \lhcb collisions, and falling in the \lhcb acceptance. The momentum distribution in \belle II is estimated by decaying \B mesons  with a momentum of 1.50 \gevc along the $z$-axis using \texttt{RapidSim}, corresponding to the $\gamma\beta=0.28$ boost of the centre-of-mass system in \belle II when operated at the $\Upsilon(4S)$ resonance~\cite{BELLE2-PUB-2018-001}. A perfect $4\pi$ angular acceptance is assumed. The generated $\D\to\KS\pi^+\pi^-$ decays are uniformly distributed in phase space. The \texttt{RapidSim} samples for \lhcb are reweighted to take the relevant time acceptance into account. This is not necessary for Belle II, as all produced \KS mesons decay in the tracking volume. 
The resulting momentum distributions for the three types of sample are shown in Fig.~\ref{fig:momentum}.

The parameter $\Delta\chi$ describes the matter-interaction effect, as detailed in Section~\ref{sub:cp_violation_and_material_interaction_of_neutral_kaons}. It depends on kaon momentum and varies along a given kaon path, as the kaon intersects detector components made of different materials. In these studies, the calculations are simplified by using a constant set of average material parameters for each experimental scenario. The average material parameters can be estimated for a given experimental scenario by considering the type and length of material traversed by a kaon in the relevant sub-detector(s). A detailed description of the calculation is given in appendix~\ref{app:detector_details}. The average value of the dimensionless parameter $r_\chi=\frac{1}{2}\frac{\Delta\chi}{\Delta\lambda}$, which governs the size of the matter regeneration effect, can be calculated for the three considered experimental scenarios, and the averages are found to satisfy $|r_\chi^\text{LL}|=2.7\times10^{-3}$, $|r_\chi^\text{DD}|=2.2\times10^{-3}$, and ${|r_\chi^\text{Belle II}|=1.0\times10^{-3}}$. 

The \lhcb detector is undergoing a significant upgrade prior to the start of the LHC Run~3. However, the material budget and geometry of the relevant sub-detectors will be similar to the sub-detectors used during Run~1~and~2~\cite{VELOUpgradeTDR,PIDUpgradeTDR}. Hence the results of this study will be valid for measurements during the upgrade phases of \lhcb, even though the detector parameters presented in this section relate to the original \lhcb detector.

\subsection{Calculation procedure} 
\label{sub:calculation_procedure}

In the numerical bias studies studies, the amplitude model for $\Dz\to\KS\pip\pim$ decays in Ref.~\cite{DalitzModel18} is taken to represent the $A_1(\spm)$ amplitude. Then $A_2(\spm)$ is obtained as described in Section~\ref{sub:cp_violation_and_material_interaction_of_neutral_kaons}. In terms of $A_1$ and $A_2$, the amplitudes $\ADorDbSL(\spm)$ can be expressed and related via Eqs.~\eqref{eq:A12toKS}~and~\eqref{eq:DDbar_relations}, and the full $\mathcal{A}^\pm_\text{S/L}(\spm)$ amplitudes calculated for a given set of input parameters $(\gamma^0, r_B^0, \delta_B^0)$.
Then Eq.~\eqref{eq:mat_time_dep} gives the kaon state as a function of time, phase-space coordinates, and the material parameter $\Delta\chi$. The neutral kaon state components, $\psi_\text{S}(t)$ and $\psi_\text{S}(t)$, are inserted into Eq.~\eqref{eq:Gamma_Bminus_with_KL}, which is integrated numerically over time and the phase-space bins of Fig.~\ref{fig:optimal_binning_scheme} to obtain the expected yields in each bin. These integrals use the experimental time acceptance that was described in Section~\ref{sub:material_parameters}. 
The signal yields depend on the momentum via the time-acceptance parameters $\tau_1$ and $\tau_2$, and because the material interaction parameter $\Delta\chi$ is momentum dependent. Therefore, the yields are averaged over the \KS momentum distributions of \lhcb and \belle II. The neutral kaon momentum in the lab frame is correlated with $m^2(\pip\pim)$, and in order to take this correlation into account in the averaging, the kaon $p$, $p_z$, and $p_T$ distributions are extracted for a number of different $m^2(\pip\pim)$ values, using the \texttt{RapidSim} samples described in Section~\ref{sub:material_parameters}.  
In order to keep the calculations manageable, the distributions of kaon $p$, $p_z$, and $p_T$ for each phase-space point are divided into 5 quantiles and the 5 medians of these quantiles are used to represent the overall distribution.

The parameters \xpm and \ypm are determined by a maximum likelihood fit to the calculated yields, using the default yield expression in Eq.~\eqref{eq:base_yields}, which ignores the presence of \CP violation and material interaction in the neutral kaon sector. The fit result and covariance matrix are interpreted in terms of the physics parameters $(\gamma, r_B, \delta_B)$ using another maximum likelihood fit~\cite{Gammacombo2016}, to allow for the extraction of the bias $\Delta\gamma = \gamma - \gamma^0$. In the fits, the $K_i$ are obtained using the definition $K_i=K_i^\text{meas}=({N^\D_i + N^\Dbar_{-i}})/({\sum_j N^\D_j +N^\Dbar_{-j}})$, in terms of the expected yields $N^\D_i$ ($N^{\Dbar}_i$) of a flavour-tagged \Dz (\Dzb) decays in bin $i$ of the \D decay phase space, calculated as described above for $r_B^0=0$. This corresponds to experimentally measuring the $K_i$ in a control channel, and takes the effect of neutral kaon \CP violation and material interaction on $K_i$ measurements into account, as well the experimental time acceptance. The $(c_i, s_i)$ are calculated using $A_1(\spm)$ and the experimental time acceptance is taken into account in this calculation as well. While a model-independent method is specifically used here to determine biases, it is expected that traditional and new unbinned methods such as those in Refs~\cite{BELLE2004,BELLE2006,BELLE2010,BABAR2005,BABAR2008,BABAR2010,LHCb-PAPER-2014-017} and Ref.~\cite{Poluektov2018}, respectively, will be similarly biased if the kaon \CP-violation and regeneration are not accounted for.

\subsection{Results} 
\label{sub:bias_results}

\begin{figure}[tbp]
    \centering
    \includegraphics{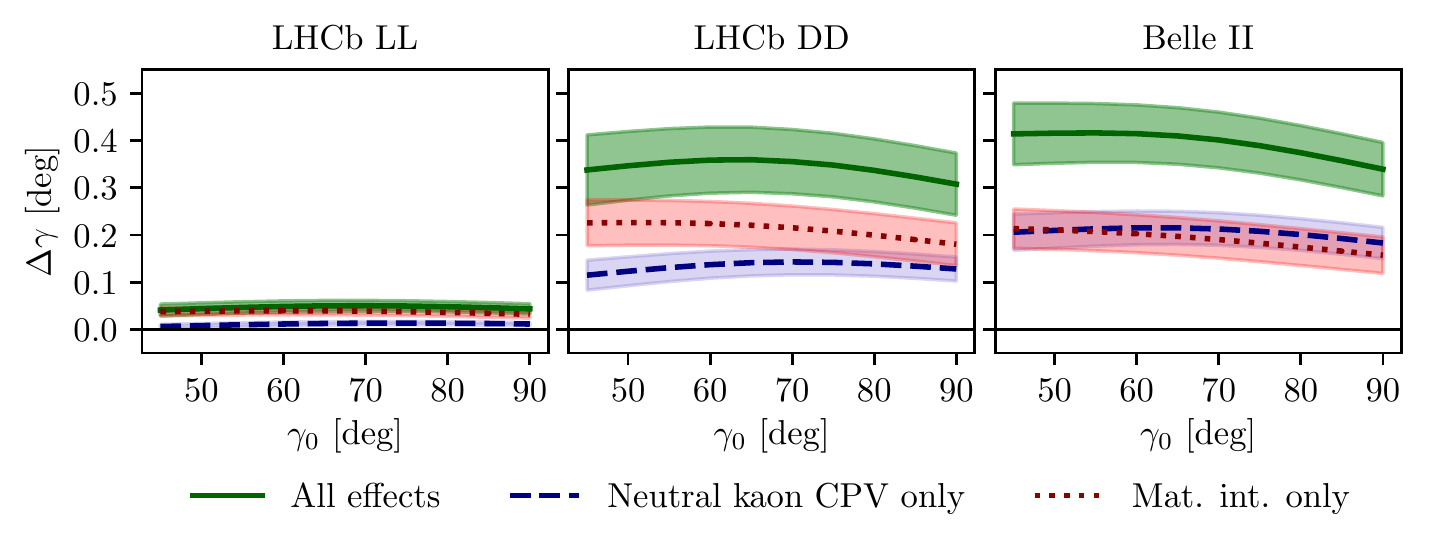}
    \caption{The bias $\Delta \gamma$ as a function of input $\gamma_0$ for (left) the LL \lhcb category, (centre) the DD \lhcb category, and (right) \belle II. The bias is calculated due to (blue, dashed line) neutral kaon \CP violation alone, (red, dotted line) material interaction alone, and (green line) both effects. The shaded region shows the estimated $1\sigma$ uncertainty band.}
    \label{fig:compare_experiments}
\end{figure}

The obtained bias $\Delta\gamma$ is shown as a function of input $\gamma^0$ for the various experimental conditions in Fig.~\ref{fig:compare_experiments}. The calculations are made using $(r_B^0, \delta_B^0)=(0.1, 130^\circ)$, approximately equal to the physics parameters relevant for $\Bpm\to\D\Kpm$ decays~\cite{HFLAV16,UTfit-UT}.  The bias does not vary significantly with $\gamma^0$ in the plotted range, which includes the world average value of direct $\gamma$ measurements as well as the values obtained in full unitarity-triangle fits~\cite{HFLAV16,UTfit-UT,CKMfitter2015}, and for all cases, the bias is found to be below $0.5^\circ$, corresponding to relative biases of about half a percent. Thus the biases are of $O(r\epsilon /r_B)$ as expected, given the arguments of Section~\ref{sec:Theoretical_background}. The contributions from the individual \KS CPV and material interaction effects are also shown. It is seen that the neutral kaon \CP violation and material interaction effects leads to approximately equal biases in all three cases. 

Given the decay-time acceptance and momentum distribution for each experimental category, the mean life time, $\langle\tau\rangle$, of the reconstructed kaons can be calculated. In terms of the \KS lifetime ${\tau_\KS=(0.895\pm0.004)\times10^{-11}\,}$s~\cite{PDG2018}, $\langle\tau_\text{LL}\rangle\simeq0.1\tau_\KS$ for the \lhcb LL category, $\langle\tau_\text{DD}\rangle\simeq0.8\tau_\KS$ for the \lhcb DD category, and at \belle II $\langle\tau_\text{Belle II}\rangle\simeq\tau_\KS$. The difference in average kaon lifetime is reflected in the observed biases, which are found to be larger in the samples with longer lived kaons. The very small effect in the LL category is to be expected because the \CP-violation effect due to \KS not being \CP-even is approximately cancelled by the \CP-violation effect arising from $\KS-\KL$ interference for kaons with decay times much smaller than $\tau_\KS$~\cite{Grossman2012}. The time dependence of the bias effect means that it can potentially be beneficial to restrict a measurement to using short-lived \KS mesons in a future scenario, where the impact of \KS\ \CP violation is comparable to the statistical precision of the measurement. For example, the bias can reduced by 40\,\% in the \belle II scenario if only \KS mesons decaying within 8\cm of the beam axis are included in the measurement. This requirement only removes 20\,\% of the signal yield, and hence only increases the statistical uncertainty of the measurement by 10\,\%.

The uncertainty bands in Fig.~\ref{fig:compare_experiments} are calculated by repeating the study while varying some of the inputs. The model dependence of the predicted biases is probed by repeating the study using two other amplitude models as input for $A_1(\spm)$ and $A_2(\spm)$: the model published in Ref.~\cite{BELLE2010} and the model included in \sc{EvtGen}\normalfont~\cite{EvtGen}. The use of different models change the predicted biases by up to $0.05^\circ$. When defining $A_2(\spm)$ in terms of $A_1(\spm)$, there is an uncertainty due to the unknown $(r_k, \delta_k)$ parameters used to describe the $\pi\pi$ resonance terms. This uncertainty is assessed by making the study with 50 different random realisations of the parameter set. The phases $\delta_k$ are sampled uniformly in the interval $[0, 2\pi]$ while the $r_k$ are sampled from a normal distribution with $\mu=\tan^2\theta_C$ and $\sigma=\mu/2$. The uncertainty is about $0.05^\circ$ across the three experiments considered. The studies are repeated while varying the time acceptances and material densities with $\pm 10\,\%$. The largest deviations in biases are found to be below $0.05^\circ$. The dependence on the handling of the momentum distribution is estimated by repeating the study using 10 and 20 quantiles to describe the momentum distributions at each point in phase space, instead of 5. The variation in the results is taken as the systematic uncertainty, and found to be below $0.01^\circ$ for all experiments. There is an additional uncertainty due to the use of simulation samples generated with \texttt{RapidSim} to describe the kaon momentum distribution, in lieu of full detector simulations. The uncertainty has not been considered here. Full detector simulation should be used if specific experimental measurements are to be corrected for the biases described in this study.

There is also an uncertainty from the use of $(\ci, \si)$ as calculated using $A_1(\spm)$. It is to be expected that the measured values $(\hat c_i, \hat s_i)$ from the \cleo collaboration differ by those calculated using $A_1^D(s_-,s_+)$ by terms of $O(\epsilon)$ due to neutral kaon \CP violation, which is not taken into account in the measurement~\cite{CLEOCISI}. These corrections can be calculated via a procedure analogous to the one used to estimate the corrections on measurements of $\gamma$ in this paper. However, as these corrections are much smaller than the experimental uncertainties in the measurement, they have not been studied further.

It is interesting to evaluate the bias obtained if the \Ki are calculated from $A_1(\spm)$, without any corrections due to neutral kaon \CP violation and material interaction. If this is done, while the full experimental time acceptance is taken into account, the biases only change by up to $0.01^\circ$, across the experiments. This is because the $O(\epsilon)$ corrections in Eq.~\eqref{eq:hat_Ki}, where the expected measured \Ki is given to lowest order in $\epsilon$ and $r_\chi$, only affect the overall normalisation. If the time acceptance \emph{is not} taken into account, biases of several degrees can occur, irrespective of the presence of neutral kaon \CP violation or material interaction effects.

\begin{figure}[tbp]
    \centering
    \includegraphics{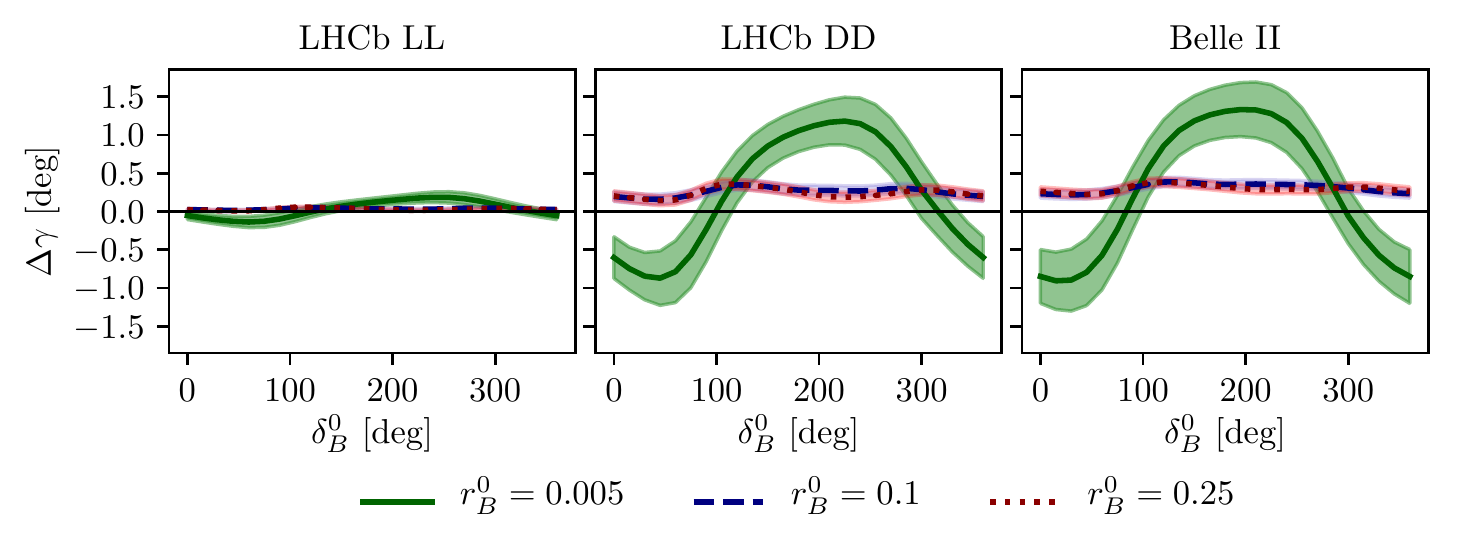}
    \caption{The bias $\Delta \gamma$ as a function of input $\delta_B$ for (left) the LL \lhcb category, (centre) the DD \lhcb category, and (right) \belle II. The bias is calculated for $\gamma=75^\circ$ and (green line) $r_B=0.005$, (blue, dashed line) $r_B=0.1$, and (red, dotted line) $r_B=0.25$. The shaded region shows the estimated $1\sigma$ uncertainty band.}
    \label{fig:delta_scan}
\end{figure}

While the $\Bpm\to\D\Kpm$ decay mode provides the best sensitivity to $\gamma$, it is also possible to measure $\gamma$ in other \B decay channels, such as $\Bpm\to\Dstar\Kpm$, $\Bpm\to\D\Kstarpm$, $\Bz\to\D\Kstarz$, and $\Bpm\to\D\pipm$. For the purpose of the study presented here, the main difference between the decay channels is that they have different values of $r_B$ and $\delta_B$. Figure~\ref{fig:delta_scan} shows $\Delta\gamma$ as a function of input $\delta_B^0$, for $\gamma^0=75^\circ$ and three different values of $r_B^0$. Aside from $r_B^0=0.1$, the results are shown for $r_B^0=0.005$, which corresponds to the expectation in $\Bpm\to\D\pipm$ decays~\cite{rDpiPaper} and $r_B^0=0.25$, which corresponds to $\Bz\to\D\Kstarz$ decays~\cite{Gammacombo2016,Gammacombo2018}. Three features are notable, namely that the biases depend on $\delta_B^0$, that the biases are large for the small $r_B^0=0.005$ case, and that the oscillation period of the $\delta_B$ dependence is different between the $r^0_B=0.005$ case and the $r_B^0\in\{0.1, 0.25\}$ cases. It is to be expected that $\Delta\gamma$ oscillates as a function of $\delta^0_B$, because $\delta_B^0$ enters the yield equations via $\cos(\delta_B^0\pm\gamma)$ and $\sin(\delta_B^0\pm\gamma)$ terms. The $r^0_B$ dependent behaviour is governed by the relative importance of different $O(r\epsilon)$ correction terms to the phase-space distribution. There are terms of both $O(r_A\epsilon)$ and $O(r_B\epsilon)$\footnote{There are similar terms of $O(r_Ar_\chi)$ and $O(r_Br_\chi)$, but as $\epsilon$ and $r_\chi$ are of the same order of magnitude, these terms can be treated completely analogously to the $O(r_A\epsilon)$ and $O(r_B\epsilon)$ terms, and have been left out of the discussion for brevity.}, which lead to expected biases of size $O(r_A\epsilon/r_B)$ and $O(r_B\epsilon/r_B)=O(\epsilon)$, respectively, cf. the discussion of Section~\ref{sub:yields_to_lowest_order}. The $O(r_A\epsilon)$ terms are independent of $\delta_B^0$, whereas the $O(r_B\epsilon)$ terms have factors of $\cos(\delta_B^0\pm\gamma)$ and $\sin(\delta_B^0\pm\gamma)$. Therefore the $O(r_A\epsilon)$ and $O(r_B\epsilon)$ terms introduce biases with different dependence on $\delta^0_B$.
In the $\Bpm\to\D\pipm$ case, the $O(r_A\epsilon)$ correction terms dominate because $r_A/r_B\simeq (0.05/0.005)=10$. This explains the relatively large bias, as $|r_A\epsilon/r_B^{D\pi}|\simeq 4\%$, and the simple dependence on $\delta^0_B$. The bias is seen to be up to {}$\pm1.5^\circ$, but only about $+0.2^\circ$ with the expected value of $\delta_B^{D\pi}\simeq300^\circ$~\cite{Gammacombo2016,rDpiPaper}. In the $r_B^0=0.1$ and $r_B^0=0.25$ cases the $O(r_B\epsilon)$ correction terms dominate, and the biases are of $O(\epsilon)$, independent of the $r_B^0$ value. Therefore both cases have biases of similar size and with similar $\delta^0_B$ dependence. While the input value of $\gamma^0=75^\circ$ was chosen for these studies, there is minimal variation in the results if another value of $\gamma^0$ in the range $[65^\circ, 85^\circ]$ is used.

The $\gamma$ measurements treated in this paper can be made using other \D-decay final states, such as ${\D\to\KS\Kp\Km}$ and $\D\to\KS\pip\pim\piz$. The biases from neutral kaon \CP violation and material interaction on measurements of $\gamma$ based the \D decay phase-space distributions should be of similar size in these decay channels, as those presented for $\D\to\KS\pip\pim$ in this paper. The impact on $\gamma$ measurements based on the phase-space-integrated yield asymmetry can be expected to be tens of degrees for the $\D\to\KS\Kp\Km$ channel, where the yield asymmetry is expected to be around 2\,\%, for the reasons explained in Section~\ref{sub:yields_to_lowest_order}. The $\D\to\KS\pip\pim\piz$ decay, however, is dominantly \CP-odd~\cite{CLEOKSpipipi0}, and the bias in measurements based on the total asymmetry is therefore expected to be $O(r_B\epsilon)$, ie. a few degrees~\cite{YuvalKsCPV}. More precise calculations of the biases would require a repeat of the study included here, with relevant amplitude models and binning schemes in place.

The studies presented here can be used to assign systematic uncertainties to measurements while the statistical uncertainties continue to dominate. As the statistical uncertainty becomes comparable with the bias effects described in this paper, the systematic uncertainty should be assigned by repeating the studies with a detailed detector simulation. This would incorporate a more accurate description of the \KS decay-time acceptance, of the full selection criteria, and the traversed material. The detailed calculations can also be used to apply a bias correction if desired.

\section{Conclusion} 
\label{sec:conclusion}

\CP violation and material interaction of neutral kaons constitute sources of bias in measurements of the CKM angle $\gamma$ based on $\Bpm\to\D(\to\KS\pip\pim)\Kpm$ decays. The relative induced bias due to these effects has been shown to be of permille level for measurements based on the difference in distributions over the \D decay phase space between signal decays originating from \Bp and \Bm mesons. However, measurements based only on the phase-space-integrated yield asymmetry between \Bp and \Bm decays, have been show to suffer biases as large as tens of degrees. The expected biases in measurements based on the phase-space distribution have been estimated for experimental conditions corresponding to \lhcb and \belle II and found to be below $0.5^\circ$, which is smaller than the precision the experiments expect to reach in this decay mode during  their lifetimes.


\acknowledgments

We thank Markus R\"ohrken for providing an implementation of the latest amplitude model, Jim Libby for his helpful consultation on the Belle II detector, and Guy Wilkinson for stimulating discussions. We are grateful for support from the Royal Society, the ERC,  the Louis-Hansen Foundation, Knud H\o jgaard's Foundation and the Augustinus Foundation.


\appendix 
\section{Details of the detector parameterisation} 
\label{app:detector_details}

To model the decay-time acceptance, requirements are placed on where the neutral kaon can decay within the detector. At LHCb the relevant detector components are the silicon vertex detector closest to the beam pipe (the VELO) and the ring-imaging Cherenkov detector which is positioned in between the VELO and the downstream trackers (the RICH1). For the LL \lhcb category, it is required that the kaon decays before reaching $z_{max}=280\mm$, corresponding to a decay where the decay products traverse at least 3 VELO segments (ignoring a number of widely spaced VELO segments placed at a distance of up to $z=750\mm$ from the interaction point)~\cite{CERN-LHCC-2003-030}. For the DD \lhcb category a decay at $z\in[280,2350]\mm$ is required, corresponding to decay between the LL cut-off and the first downstream tracking station~\cite{LHCb-2003-140}. For \belle II, it is assumed that the \KS reconstruction is similar to the \belle\ \KS reconstruction, which is based on a neural network and reconstructs \KS decays for which the decay product leave tracks in the drift chamber only, as well as decays with decay products that leave tracks in both the drift chamber and silicon vertex detectors~\cite{BelleKSPaper,BelleKSThesis}. Therefore, the \KS decay is required to be within $r_{max}=1130\mm$ of the beam axis, corresponding to a decay within the outer radius of the drift-chamber~\cite{BELLE2-PUB-2018-001}. In practice, most of the kaons decay inside the silicon vertex detector, the outermost layer of which is at $r=140\mm$~\cite{BELLE2-PUB-2018-001}, so requiring a decay before 1130\mm is essentially equivalent to having no time cut-off.

The parameter $\Delta\chi$ that describes the material interaction in Eq.~\eqref{eq:mat_time_dep} is related to the forward scattering amplitude $f$ $(\bar f)$ of \Kz (\Kzb) mesons in a given material~\cite{Good1957,Fetscher1996}
\begin{align}\label{eq:mat_delta_chi}
    \Delta \chi = - \frac{2\pi \mathcal{N}}{m_K}(f-\bar f) = - \frac{2\pi (N_A \rho/A)}{m_K}(f-\bar f),
\end{align}
where $\mathcal{N}=N_A\rho/A$ is the scattering centre density of the material, $m_K$ is the mass of the kaon state,  $A$ and $\rho$ are the nucleon number and density of the material, and $N_A$ is Avogadro's number.
Measurements made for a range of nuclei~\cite{Gsponer1979} show that in the momentum range $p_K\in [20, 140]\gevc$
\begin{align} \label{eq:f_p_dep}
    \left|\frac{f-\bar f}{p_K}\right| = 2.23 \frac{A^{0.758}} {p_K^{0.614} (\gevc)} \text{ mb}, \quad \arg [f - \bar f] = -\frac{\pi}{2}\left(2-0.614\right),
\end{align}
where the phase of $\Delta f$ is determined via a phase-power relation~\cite{Briere1995}. In the numerical studies presented here, Eq.~\eqref{eq:f_p_dep} is also used for the low momentum neutral kaons in the \belle II calculations, as a more detailed modelling of the low momentum $\Delta\chi$ based on Ref.~\cite{Ko2011} is found to yield very similar results. The scattering centre density $\mathcal{N}$ is approximated as being constant, equal to the average density along a neutral kaon path due to its intersection with different detector segments. This average is estimated using the simplifying assumption that the total detector material budget is due to silicon. In practice, $\mathcal{N}=N_A\rho/A$ is calculated using $A=28$ and $\rho = f^\text{Si}\rho^\text{Si}$, where $f^\text{Si}<1$ is the average fraction of a neutral kaon path length that is inside detector material, estimated via the known dimensions of the detector, the average nuclear interaction length seen by a track traversing it, and the nuclear interaction length of silicon $\lambda_I^\text{Si}=465.2\mm$~\cite{PDG2018}. To calculate the average material density for the LL \lhcb category, the full length of the VELO, $L_{VELO}=750\mm$, is used, as the average $\Delta x/\lambda_I=3.8\,\%$ is given for particles traversing the whole detector in Ref.~\cite{CERN-LHCC-2003-030}. The calculation yields ${f^\text{Si}_\text{LL}=(\Delta x/\lambda_I)\times \lambda_I^\text{Si}/L_{VELO} =2.4\,\%.}$ 

The corresponding number in the DD category, in which the neutral kaons traverse both the VELO and the RICH1 detector, is obtained by assuming that the space between $z=750\mm$ and $z=2350\mm$ is taken by the RICH1, in which the average nuclear interaction length seen is $\Delta x/\lambda_I=2.6\,\%$~\cite{LHCb-2004-121}. Averaging the density between the VELO and RICH1 detectors yields $f^\text{Si}_\text{DD}=1.6$ \% (the average is weighted with a factor, ${\exp[-t/\tau_\KS]=\exp[-z m_K/(\tau_\KS p_\KS)]}$, using the mean \KS momentum, to take into account that more \KS mesons decay earlier, where the density is higher, rather than later). 

In \belle II the particle travels approximately 6 \% of a radiation length in the beam pipe and vertex detectors~\cite{Waleed2017}, which stretch to $r=140$ mm~\cite{BELLE2-PUB-2018-001}, and another 6 \% of a radiation length in the drift chamber~\cite{Waleed2017} (using the approximate material budget in the central region $\theta\in[35^\circ, 120^\circ]$), which stretches to $r=1130$ mm~\cite{BELLE2-PUB-2018-001}. A calculation analogous to those above, but using the radiation length of silicon $X_0^\text{Si}=93.7\mm$~\cite{PDG2018}, yields $f^\text{Si}_\text{Belle II}=3.8$\,\%. As for the \lhcb calculations, the full radial length of the detector segments are used because the material budget is given for tracks traversing the whole detector, and the average is calculated with a weight factor of $\exp[-z m_K/(\tau_\KS p_\KS)]$.

The average value of $r_\chi=\frac{1}{2}\frac{\Delta\chi}{\Delta\lambda}$, which governs the size of the matter regeneration effect, can be calculated for the three considered experimental scenarios and satisfy $|r_\chi^\text{LL}|=2.7\times10^{-3}$, $|r_\chi^\text{DD}|=2.2\times10^{-3}$, and ${|r_\chi^\text{Belle II}|=1.0\times10^{-3}}$.

\bibliography{main}{}
\bibliographystyle{JHEPmod}

\end{document}